\documentclass[preprint]{elsarticle}
\usepackage[utf8]{inputenc}
\usepackage{url}
\usepackage{amsmath}

\makeatletter
\def\ps@pprintTitle{%
   \let\@oddhead\@empty
   \let\@evenhead\@empty
   \def\@oddfoot{\reset@font\hfil\thepage\hfil}
   \let\@evenfoot\@oddfoot
}
\makeatother

\begin{document}
\begin{frontmatter}

\title{New Search for Mirror Neutrons at HFIR}

\author[ornl]{L.~J.~Broussard} 
\author[ornl]{K.~M.~Bailey}    
\author[ornl]{W.~B.~Bailey}    
\author[utk]{J.~L.~Barrow} 
\author[utk]{B.~Chance} 
\author[uky]{C.~Crawford} 
\author[ornl]{L.~Crow}     
\author[ornl]{L.~Debeer-Schmitt} 
\author[utk]{N.~Fomin} 
\author[ornl,utk]{M.~Frost} 
\author[ornl]{A.~Galindo-Uribarri} 
\author[ornl]{F.~X.~Gallmeier} 
\author[utk]{L.~Heilbronn} 
\author[ornl]{E.~B.~Iverson} 
\author[utk]{Y.~Kamyshkov} 
\author[indiana]{C.-Y.~Liu} 
\author[wku]{I.~Novikov} 
\author[ornl]{S.~I.~Penttil{\"a}} 
\author[utk]{A.~Ruggles} 
\author[utk]{B.~Rybolt} 
\author[indiana]{M.~Snow} 
\author[utk]{L.~Townsend} 
\author[utk]{L.~J.~Varriano} 
\author[utk]{S.~Vavra} 
\author[ncsu]{A.~R.~Young} 

\address[ornl]{Oak Ridge National Laboratory, Oak Ridge, TN 37831  USA}
\address[utk]{University of Tennessee, Knoxville, TN 37996  USA}
\address[indiana]{Indiana University, Bloomington, IN  47405  USA}
\address[uky]{University of Kentucky, Lexington, KY 40506  USA}
\address[wku]{Western Kentucky University, Bowling Green, KY 42101  USA}
\address[ncsu]{North Carolina State University, Raleigh, NC 27695  USA}

\date{September 2017}

\begin{abstract}
The theory of mirror matter predicts a hidden sector made up of a copy of the Standard Model particles and interactions but with opposite parity. If mirror matter interacts with ordinary matter, there could be experimentally accessible implications in the form of neutral particle oscillations.  Direct searches for neutron oscillations into mirror neutrons in a controlled magnetic field have previously been performed using ultracold neutrons in storage/disappearance measurements, with some inconclusive results consistent with characteristic oscillation time of $\tau$$\sim$10~s. Here we describe a proposed disappearance and regeneration experiment in which the neutron oscillates to and from a mirror neutron state. An experiment performed using the existing General Purpose-Small Angle Neutron Scattering instrument at the High Flux Isotope Reactor at Oak Ridge National Laboratory could have the sensitivity to exclude up to $\tau$$<$15~s in 1 week of beamtime and at low cost.

\end{abstract}

\end{frontmatter}

\section{Introduction}

A strong foundation of evidence has been built over the past several decades which indicates that 85\% of the matter in the universe is unknown to us~\cite{Patrignani:2016xqp}. The evidence for Dark Matter comes from many diverse astronomical sources including rotational curves of galaxies, weak and strong lensing, galaxy collisions such as the Bullet Cluster, and the cosmic microwave background; however, to date it is based solely on astrophysical signatures of its gravitational influence. Determining the particle nature of dark matter has for years been one of the highest priorities of the particle physics community, but there remains no conclusive evidence to its origin.  A significant focus of these searches has been the well-motivated Weakly Interacting Massive Particles~\cite{Tan:2016zwf,Akerib:2016vxi}, but the diminishing parameter space above the neutrino background provides strong motivation to consider alternate theories. The 2014 Report of the Particle Physics Project Prioritization Panel (P5) stressed the importance of considering ``every feasible avenue,'' culminating in a recent community workshop to explore the science case for small-scale projects which can explore beyond this parameter space~\cite{Battaglieri:2017aum}. 

The possibility of mirror matter as a type of hidden sector dark matter candidate has been considered for decades~\cite{Kobzarev:1966qya, Blinnikov:1982eh,Khlopov:1989fj,Foot:1991bp,Hodges:1993yb} (see also reviews~\cite{Berezhiani:2005ek,Okun:2006eb,Foot:2014mia}).  Mirror matter manifests as a perfect copy of Standard Model particles and interactions such that parity and time reversal are exact symmetries, and interacts very weakly with our known universe, primarily gravitationally.  It is a type of asymmetric, self-interacting, baryonic dark matter.   To meet constraints from Big Bang nucleosynthesis and the effective number of light neutrino species, the mirror universe could not be identical in terms of cosmological evolution, and should instead have a lower temperature, and therefore be helium dominated, with a larger baryon asymmetry~\cite{Berezhiani:2000gw}. Like ordinary matter, mirror matter should dissipate energy at too high a rate for halos to form; the formation of disk galaxies may be avoided by a more rapid stellar evolution~\cite{Berezhiani:2005vv}, or the energy dissipation rate may be compensated via an energy injection from supernovae due to kinetic mixing between the two sectors~\cite{Foot:2004wz}.

\section{Neutron Oscillations}

If the mirror universe only interacts gravitationally with the ordinary universe, it is not of much interest as there are no testable consequences that could be performed by particle physicists. Instead we consider the possibility of mixing between the two universes which could manifest as neutral particle oscillations. Photons, neutrinos, neutral pions and kaons, and neutrons are good candidates for consideration. Neutron oscillations ($n\rightarrow n'$) in particular are interesting due to the implications of a new baryon number violating process. The phenomenology of mirror neutron oscillations was considered with the realization that strong limits on the possibility of rather fast oscillation times did not yet exist~\cite{Berezhiani:2005hv}, and a more detailed treatment followed including the consideration of a small but nonzero mirror magnetic fields, which could originate from mirror matter accumulated in the Earth~\cite{Berezhiani:2008bc}. 

In this model, the Hamiltonian of the free neutron in the presence of nonzero ordinary and mirror magnetic fields $B$ and $B'$ respectively is of the form
\begin{equation}
   H=
  \left( {\begin{array}{cc}
   \mu \vec{B}\cdot\vec{\sigma} & \tau^{-1} \\
   \tau^{-1}  & \mu' \vec{B'}\cdot\vec{\sigma'}  \\
  \end{array} } \right)
\end{equation}
where $\vec{\sigma}$ is the neutron spin, and $\tau$ is the characteristic $n\rightarrow n'$ oscillation time, $\mu$ is the neutron magnetic moment, and $\mu = \mu'$ as a consistent assumption of the mirror matter model. The $n\rightarrow n'$ oscillation probability versus free flight time $t$ for unpolarized neutrons is~\cite{Berezhiani:2012rq}
\begin{align}
P(t) &= \frac{sin^2\left[(\omega-\omega')t\right]}{2\tau^2(\omega-\omega')^2} +  \frac{sin^2\left[(\omega+\omega')t\right]}{2\tau^2(\omega+\omega')^2} \nonumber \\
&+ \left(cos\beta\right) \left( \frac{sin^2\left[(\omega-\omega')t\right]}{2\tau^2(\omega-\omega')^2} -  \frac{sin^2\left[(\omega+\omega')t\right]}{2\tau^2(\omega+\omega')^2}  \right) 
\end{align}
where $\omega=\frac{1}{2}|\mu B|$ and $\omega'=\frac{1}{2}|\mu' B'|$ and $\beta$ represents the angle between $\vec{B}$ and $\vec{B'}$. The $n\rightarrow n'$ oscillation probability scales with the free neutron flight time as $\sim \frac{t^2}{\tau^2}$ when $\vec{B}\approx\vec{B'}$ and exhibits a resonance behavior when the difference in the magnitudes of the magnetic fields is small. The dependence on the magnetic field direction is contained in $cos \beta$: the probability is maximal when the fields are aligned, but the components with the resonance condition are cancelled when the fields are anti-aligned.

Previous published limits on $n\rightarrow n'$ oscillations~\cite{Patrignani:2016xqp} have used ultracold neutrons: neutrons which have energy $<\sim$300~neV such that they can be totally internally reflected by the Fermi potential of some materials and thus be stored in material bottles. The most stringent limit available, $\tau<414$~s (90 \% C.L.), was obtained in an experiment which examined the storage time of ultracold neutrons in a material bottle with and without the presence of an external magnetic field~\cite{Serebrov:2008hw}; however, this limit assumes $\vec{B'}=0$, or no mirror magnetic field present at the Earth. When reanalyzed to consider the possibility $\vec{B'}\neq0$, an anomalous signal consistent with $\tau\sim10$~s and $B'\sim100$~mG was observed with 5~$\sigma$ significance~\cite{Berezhiani:2012rq}. Another ultracold neutron storage experiment also scanned the magnetic field up to $\pm$125~mG and excluded $\tau<12$~s (95\% C.L.) for this range~\cite{Altarev:2009tg}. The sensitivity of this experiment was limited primarily by the step size in the magnetic field scan, 25~mG, which was much larger than the resonance width.

Potential systematic effects which can induce unmonitored changes in the storage time of ultracold neutrons could include wall-loss probabilities which depend on magnetic field or spectral evolution of the ultracold neutron population on a timescale shorter than the field scan time. 
An independent approach with different systematic considerations is needed to resolve these controversial results.  A cold neutron disappearance and regeneration experiment could give a clear, unambiguous indication of mirror neutron oscillation. 

\section{Cold Neutron Regeneration}

As in many other measurements in fundamental neutron physics, it is useful to consider the unique advantages available to experiments using either cold or ultracold neutrons. A particular advantage of ultracold neutron experiments is the potentially long storage time of the neutrons, enabling a relatively compact experiment in which each slow-moving neutron undergoes many thousands of ``bounces'' which can sample the oscillation probability. Alternatively, cold neutrons are produced with fluxes of many orders of magnitude higher than ultracold neutron sources. Therefore, cold neutrons offer an attractive opportunity to search not only for an anomalous disappearance, but also the regeneration of neutrons from the mirror state.

A search for the disappearance and regeneration of neutrons using cold neutrons has been attempted previously~\cite{Schmidt2007}, and more recently the technique has been described in detail which includes considerations for a local mirror magnetic field~\cite{Berezhiani:2017azg}. The cold neutron approach includes two separate stages which could be developed and run either independently or simultaneously for a more powerful check against a systematic-induced anomaly. The approach is depicted in Figure~1.  

The first stage, the disappearance search, involves a precise measurement of the controlled disappearance of a large beam flux of cold neutrons. A magnetic field control system provides a uniform magnetic field over some length of the beamline which can be ``tuned'' to match the unknown mirror magnetic field and achieve the resonance for the neutron oscillation probability.  A signal for disappearance into a mirror neutron state is given by a reduction in the flux for a particular magnitude of the magnetic field. The neutron flux can be monitored upstream of the oscillation section by a low efficiency neutron counter. 

For a regeneration search, the neutron counter at the first stage beamline exit is backed by a beam stop to significantly attenuate the number of neutrons that can enter the second stage of the experiment. Only mirror neutrons, whose interactions with ordinary matter are negligible, can pass through unhindered. The second stage consists of a larger diameter vacuum vessel with similar requirements for magnetic field control to enable the oscillation of mirror neutrons back into ordinary neutrons.  A low background, large area detector at the end of the second stage detects the regenerated neutrons.

\begin{figure}
\centering
\label{fig:overview}
\includegraphics[width=\textwidth]{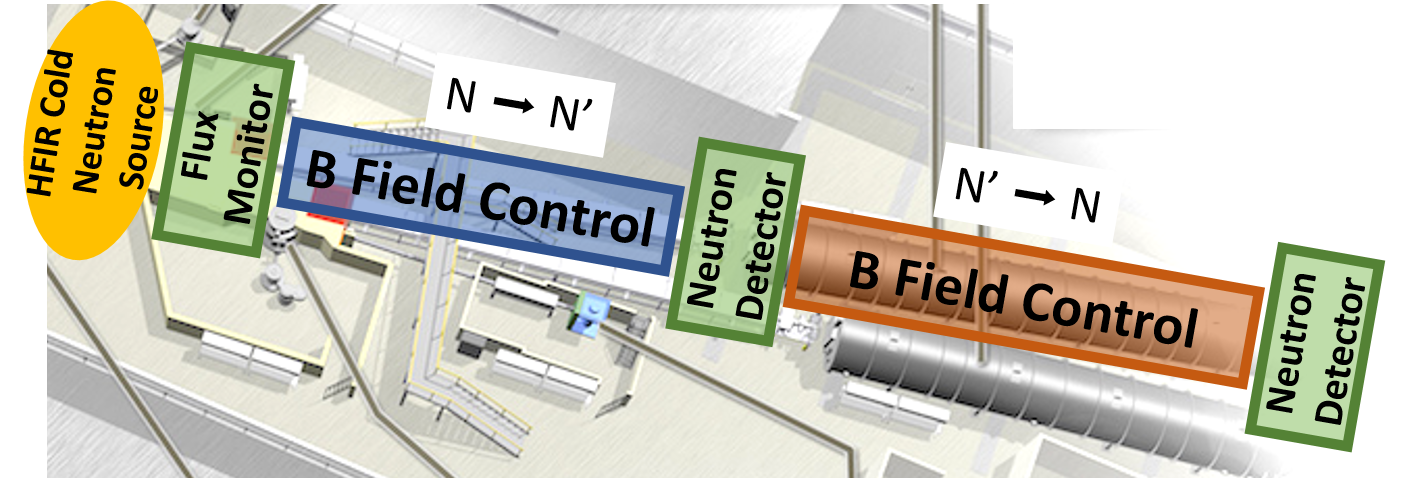}
\caption{Cold neutron disappearance~$\times$~regeneration experiment at the HFIR reactor, described in the text (color online).}
\end{figure}

\section{Mirror Neutron Oscillation Search at HFIR}

The experiment described in reference~\cite{Berezhiani:2017azg} assumes development of two 15~m long, large area beamlines as well as an up to 0.5$\times$0.5~m$^2$ area final neutron counter, to be staged at the Spallation Neutron Source (SNS) at Oak Ridge National Laboratory (ORNL); this could impose a limit on the oscillation time of a few tens of seconds, depending on the configuration. This experiment would not only require the development of vacuum flight tubes with corresponding magnetic field control system and detection system, but also the availability of an unused beam port, a requirement that represents a nontrivial effort. An experiment which can fully explore the parameter space of the previous ultracold neutron searches can be performed using an existing cold neutron beamline such as at the High Flux Isotope Reactor (HFIR) at ORNL~\cite{CROW2011S71} or the National Institute of Standards and Technology Center for Neutron Research (NIST NCNR)~\cite{DEWEY2005213}, and could produce results in a few years with modest investments.

The General Purpose Small Angle Neutron Scattering (GP-SANS) CG-2 instrument at HFIR may be an ideal candidate for developing such an experiment. The disappearance stage of the experiment will be installed in the 16~m long vacuum beamline section currently used for collimation of the neutron beam. The neutron flux was simulated using the MCNP neutron transport package and using the acceptance diagram method~(Fig.~2)~\cite{MOON}. The total neutron intensity through the 4$\times$4~cm$^2$ opening in the shielding upstream of the beamline is 2$\times10^{10}$~n/s.  The simulated spectrum peaks at a wavelength of 4~\AA ~ and the standard deviation of the beam divergence is 0.21$^\circ$ at 4~\AA ~and increases to 0.28$^\circ$ above 7~\AA. The collimation section of the beamline is scheduled to undergo an upgrade in 2018 to improve guide handling for SANS applications and to remove magnetic material to enable a polarized neutron beam. This operation will be advantageous for a mirror neutron search to better accommodate the strict magnetic field homogeneity requirements.

\begin{figure}
\centering
\label{fig:spec}
\includegraphics[width=0.75\textwidth]{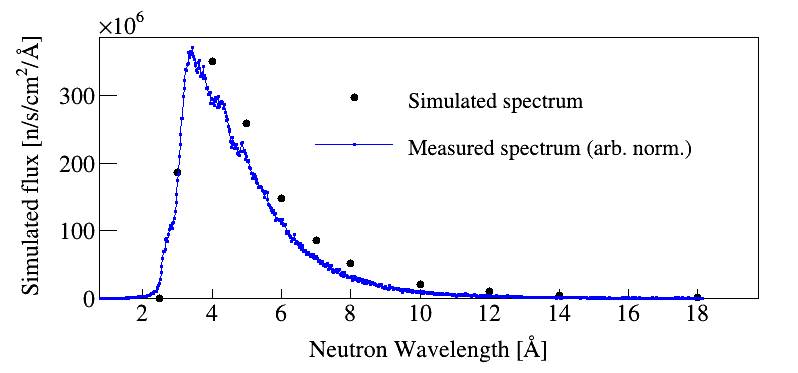}
\caption{Comparison of the shape of the simulated and measured neutron flux (arbitrary normalization) at GP-SANS used in determining sensitivity limits.  The spectrum was measured without a velocity selector and with no beam guides through the collimation section.  The measured spectrum y-axis is scaled to roughly match the simulated flux (color online).}
\end{figure}

A disappearance search using cold neutrons is in some respects significantly more challenging than using ultracold neutrons, as it involves a very high precision determination of the neutron transmission: an oscillation time of $\tau=$15~s would result in a change in flux at the few 10$^{-7}$ level. Variations in the neutron spectrum can potentially result in second-order effects that can rise to this level, which can sample, for example, polarization dependent reflectivity of the cold neutron guide or energy-dependence and spatial nonuniformities in the detection efficiency. These effects can be suppressed by scanning the magnetic field on a timescale that is faster than the variation in the neutron total flux and spectral shape.  Systematic variations can be studied by applying larger magnetic fields to exacerbate field-dependent effects or increasing the time per scanning sequence, or reducing the length of the field control region to confirm the $\frac{t^2}{\tau^2}$ dependence in the event of an observed anomalous disappearance.

The disappearance detector will use a design implemented by other sensitive cold neutron experiments studying the hadronic weak interaction which require highly accurate determinations of cold neutron transmission, such as NPDGamma~\cite{Fry:2016esm}, the $n-^3$He experiment~\cite{Fomin:2016dpo}, and the $n-^4$He spin rotation experiment~\cite{Snow:2011zza}. The detector will be a current-mode, parallel plate $^3$He ion chamber, which detects neutrons that undergo the $^3$He($n$,$p$)$^3$H reaction~\cite{PENN2001332,doi:10.1063/1.4919412}. This type of detector produces a large signal with a well-defined amplitude independent of the incident neutron energy prior to capture on the $^3$He, and is nearly insensitive to gamma radiation. The detector charge collection plates can be segmented into quadrants to suppress common-mode noise due to nonstatistical fluctuations in the reactor intensity, and can also reduce sensitivity to spatial fluctuations. This technique can be used to achieve a counting sensitivity near the statistical limit, of about 1.1$\sqrt{N}$.

The GP-SANS detection chamber represents an especially attractive opportunity for staging a regeneration search.  The detection chamber is a 20~m long, 2.5~m diameter stainless steel vacuum chamber, coated with cadmium to reduce neutron backgrounds. The $1\times1$~m$^2$ area neutron detector consists of two overlapping rows of 192 individual cylindrical $^3$He gas chambers~\cite{BERRY2012179}. The large area of the detector accommodates the needs of SANS applications and well exceeds the acceptance required for a mirror neutron regeneration search. The detection efficiency is $>90\%$ at a neutron wavelength of 5~\AA, and it has a position resolution of 5~mm~$\times$~5~mm. The regeneration signal to background can be optimized using position cuts to limit the detector area. The background count rate of the detector with the reactor off was measured to be about $2\times10^{-4}$~cps/cm$^2$ or about 2~cps over the full detector area, expected to be primarily due to cosmogenic neutrons which are moderated in the concrete floor. Further reduction may be possible with additional shielding. With the reactor on and the beamline shutter in place, but no beam stop in front of the open window of the detector chamber, there is a modest increase in background rate at the center of the detector due to punch-through. The neutron detector can be positioned between 2--20~m along the length of the chamber, which can enable a suite of systematic studies, or a more detailed study of the oscillation probability versus free flight time.

An important challenge for disappearance and regeneration searches is the implementation of the magnetic field control system. To render inhomogeneities in the magnetic field negligible, ideally the magnetic field would need to be determined to better than a few mG. The spatial and temporal nonuniformity of the current (pre-upgrade/somewhat magnetic) collimation section of beamline was studied using an array of HMC5883L magnetometer boards\footnote{https://www.adafruit.com/product/1746}, read out by an Arduino Uno and Raspberry Pi 3. The typical temporal variation (FWHM) was measured to be up to 10~mG, limited by noise in the system. Short duration changes in the field as large as 1~G were observed due to changing experimental conditions in the GP-SANS or neighboring beamlines; these periods could be vetoed during data-taking. The magnitude of the magnetic field varies by up to 500~mG along most of the beamline, but increases significantly near the beamline exit. In particular, a steel column near the beamline exit can become highly magnetized by use of a superconducting magnet during some SANS experiments; if cancellation is not successful, the usable length of the beamline could be shortened slightly. Idealized COMSOL simulations indicate that a set of solenoidal and cosine theta coils surrounded by a sheet of Mu-metal should be sufficient to provide 2~mG uniformity for a cold neutron guide diameter of about 20~cm in the disappearance section. The magnetic field control system for the larger-diameter vacuum vessel used as the regeneration section will implement control coils outside the vessel.  Design of the multi-coil system using feedback from calibrated magnetometers inside the vessels is under development.

The sensitivity of a disappearance and regeneration search at GP-SANS was investigated using a Monte Carlo simulation. The neutron flux was taken as simulated and the beam guides were assumed to be sufficiently large that the average free flight of the neutrons spanned 14~m and 20~m in the disappearance and regeneration regions, respectively. Due to the conflicting optimizations required for disappearance and regeneration, greater sensitivity can be obtained by conducting either a disappearance or a regeneration search. The neutron transmission determination for disappearance was assumed to be nearly statistics limited (1.1$\sqrt N$) with a 30$\%$ flux monitoring detector. A rough (1\% level) determination of the total neutron flux will suffice for the regeneration search. The accepted area of the regeneration neutron detector was $\sim1500$~cm$^2$ to reduce the total background to 0.3~cps. The optimal step size for magnetic field control was taken as 10~mG for disappearance and 5~mG for regeneration-focused experiments. The disappearance search can be performed by changing the magnetic field along one axis (worst case $\beta=90^\circ$), and a 2.5$\times$ improvement in sensitivity is obtained in a regeneration search by performing a scan over four directions (worst case $\beta=60^\circ$).

An experimental run consists of a sequence of rapid measurements at different magnetic field values. The measured neutron counts at each field value is summed to produce a resonance curve.  The resulting curve is then compared to the simulated expected signals over the parameter space of neutron oscillation time and magnetic field in order to extract a limit. An example of a simplified sequence in which only one axis is scanned is depicted in Fig.~3.  A positive signal for an oscillation time of $\tau$ = 14~s after 7 days beamtime each for disappearance and regeneration (optimized as described) is shown to illustrate the sensitivity to magnetic field control, signal to background, and flux normalization requirements. Either a disappearance or regeneration study can have the sensitivity to exclude up to 15~s in the range $-125$~mG~$<B'<$~125~mG (90\% C.L.) in less than 7 days beamtime. The short beamtime requirement is highly desirable as GP-SANS is a heavily subscribed Basic Energy Sciences instrument.  In the event of an anomalous detection, a compromised optimization of the beam can be used to perform both stages simultaneously, requiring several (not necessarily continuous) weeks of beamtime for similar sensitivity. A powerful check against systematics which can introduce a false anomaly is the requirement that any such anomaly must be present simultaneously in both stages only when both of the magnetic field control systems are tuned to the same field.

\begin{figure}
\centering
\includegraphics[width=\textwidth]{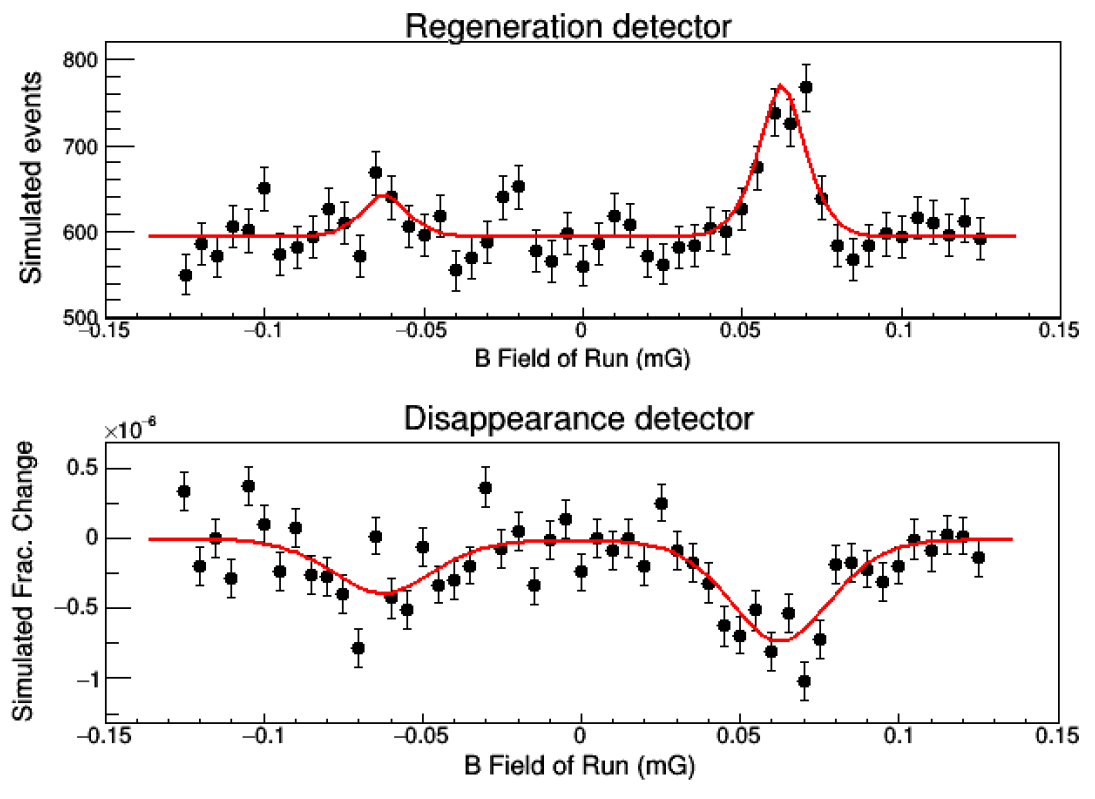}
\label{fig:sig}
\caption{Example simulated resonance curve with positive signal for $\tau=14$~s neutron oscillation time at GP-SANS at HFIR (color online).}
\end{figure}

\section{Conclusions}

Neutron oscillations into invisible mirror partners represent an experimentally accessible prediction of mirror matter, a dark matter candidate. We have described an experimental approach that is complementary to previous searches using ultracold neutrons, and which can fully explore the parameter space of $-125$~mG~$<B'<$~125~mG and $\tau<15$~s in a short beamtime of 1 week, and additional beamtime could be used to explore larger magnetic fields. The experiment will be staged at the existing GP-SANS instrument after the 2018 upgrade and will require only modest costs to implement a disappearance beamline, detection system, and magnetic field control systems, with minimal impact to the GP-SANS program. 

\section{Acknowledgements}

This research was sponsored by the Laboratory Directed Research and Development Program [project 8215] of Oak Ridge National Laboratory, managed by UT-Battelle, LLC, for the U. S. Department of Energy, and was supported in part by the U.S. Department of Energy, Office of Science, Office of High Energy Physics [contract {DE-SC0014558}] and Office of Nuclear Physics [contracts {DE-AC05-00OR2272}, {DE-SC0014622}, {DE-FG02-97ER41042}, and {DEFG02-03ER41\\258}], and in part by the National Science Foundation [contract PHY1615153].


\bibliography{DPF2017ConfProc}
\biboptions{sort&compress}

\end{document}